\documentclass[12pt]{iopart}

\usepackage{iopams}  

\usepackage{graphics} 
\usepackage[pdftex]{graphicx}  
\usepackage{amsfonts}  
\usepackage{amssymb}
\usepackage{mathrsfs}
\usepackage{subfigure}
\usepackage{color}

\renewcommand{\phi}{\varphi}
\renewcommand{\>}{\right \rangle}

\newcommand{\ket}[1]{\left |#1\>}

\newcommand{\aver}[1]{\left \langle #1\right \rangle}

\newcommand{\be}{\begin{equation}}
\newcommand{\ee}{\end{equation}}
\newcommand{\bea}{\begin{eqnarray}}
\newcommand{\eea}{\end{eqnarray}}

\begin{document}

\title[Frequency response of an atomic resonance driven by weak FEL fluctuating pulses]{Frequency response of an atomic resonance driven by weak free-electron-laser fluctuating pulses}

\author{G M Nikolopoulos$^1$ and P Lambropoulos$^{1,2}$}

\address{$^1$Institute of Electronic Structure \& Laser, FORTH, P.O.Box 1527, GR-71110 Heraklion, Greece}
\address{$^2$Department of Physics, University of Crete, P.O. Box 2208, GR-71003 Heraklion, Crete, Greece}

\ead{nikolg@iesl.forth.gr}

\begin{abstract}
Motivated by recent experiments pertaining to the interaction of weak SASE-FEL pulses with atoms and molecules, we investigate the conditions under which such interactions can be described in the framework of a simple phase-diffusion model with decorrelated atom-field dynamics. The nature of the fluctuations that are inevitably present in SASE-FEL pulses is shown to 
play a pivotal role in the success of the decorrelation. Our analysis is performed in connection with specific recent experimental 
results from FLASH in the soft X-ray regime.
\end{abstract}

\submitto{\JPB: Special Issue on 'Frontiers of FEL science'}
\maketitle

\section{Introduction}
\label{sec1}
A  ubiquitous resonant transition occurring in many contexts, involves the coupling of a ground to an
excited state by electromagnetic (EM) radiation. Aside from its role in traditional spectroscopy, in the context of laser-matter interactions, it often represents the first step in a more complex process, such as double resonance (DR), Electromagnetically Induced Transparency (EIT), etc., where it may also play the role of a probe \cite{Scully}. A basic condition for the latter is the weakness of the radiation-electron coupling, by which we mean that the corresponding Rabi frequency should be much smaller than the decay rate of the excited state. That decay could be radiative, due to spontaneous emission, autoionization, or even ionization due to the same or a second radiation source.

Under these conditions, the conventional wisdom is that the only aspect of the exciting radiation that matters  is its bandwidth and possibly the particular form of the line-shape. Excluding the special case of a Fourier limited pulse, bandwidth and line-shape depend on the stochastic properties of the radiation, which reflect the nature of the source, i.e. the processes that produced the radiation. Thus a source of thermal radiation is known to be chaotic, whose amplitude is represented by a complex Gaussian variable and well defined correlation functions \cite{Loud}. On the other hand, an ideal single-mode CW laser may to a reasonable approximation be modelled by a coherent state with constant amplitude, whose phase undergoes diffusion (a random walk from 0 to $2\pi$) \cite{Scully}. The line-shape of a source, with constant amplitude and phase fluctuations, can be rigorously modelled by a Lorentzian \cite{AgaPRA70, AgaPRA78, GeoPRA78, GeoPRA79, AgoJPB78, ZolPRA79}. A more far reaching consequence, however, stems from an equally rigorous property of the stochastic density matrix differential equations describing the interaction; namely, the decorrelation of the fluctuations of the populations from those of the Rabi frequency \cite{AgaPRA70, AgaPRA78, GeoPRA78, GeoPRA79, AgoJPB78, ZolPRA79}. This property, discussed in more detail later on, leads to considerable analytic simplification, often allowing for analytic solutions.

Several years ago, it was shown that, as long as the Rabi frequency is small, in the above sense, the decorrelation represents an excellent approximation, even in the presence of amplitude (intensity) fluctuations \cite{GeoPRA79, AgoJPB78, ZolPRA79}. In view of the ensuing analytic simplicity, this can be a very useful approximation, provided its range of validity is well defined. But aside from some applications many years ago, there has not been a systematic study of the underlying theory, which in those early days was based on the assumption of a stationary radiation source \cite{Good}. With laser sources of long pulse duration (more than a few 
ps), that assumption may have been reasonable. Present day strong lasers, however, including the short 
wavelength SASE Free Electron Lasers (FELs) which are our main interest here, are of much shorter pulse 
duration, exhibiting at the same time strong intensity fluctuations. As a consequence, stationarity can
no longer be assumed a priori. Moreover, depending on the conditions of a particular experiment, we
have the simultaneous presence of the Fourier as well as the stochastic bandwidth whose proper modelling
becomes a necessity and must be compatible with the extensive literature on the subject. 

Our purpose in this paper is to present a consistent theory, which re-evaluates the previous know how and
assumptions, in the context of sources with the properties of the short wavelength FEL. Part of our 
motivation stems from our recent experience with specific experimental results in the soft X-ray range \cite{MazJPB12}, 
one example of which is discussed in section 3. After a self-contained overview of the properties and
modelling of chaotic SASE-FEL pulses in section 2, we present a detailed treatment of the excitation
of an  Auger resonance by FEL radiation. Issues such as the effect of particular line-shapes, the
possibility of analytic solutions, and the degree of validity of the above mentioned decorrelation
are discussed in considerable detail, while our conclusions are summarized in section 4.

\section{Simulation of chaotic SASE-FEL Pulses}
\label{sec2}
The statistical properties of  light pulses emitted by a SASE-FEL depend crucially on the regime of operation \cite{SalSchYur}. 
Throughout this work, we consider light pulses produced from a SASE-FEL operating in the regime of exponential growth (linear regime). In this case it has been shown, that the pulses exhibit the properties of the so-called chaotic polarized  light \cite{SalSchYur,Kri06,Ack07}.  
The chaotic light is a fundamental concept of quantum optics, with the discussion usually limited to stationary and ergodic thermal sources (e.g., see \cite{Loud}). 
In contrast to such a type of sources, SASE-FELs are not continuous sources, and they produce random light pulses which exhibit 
spikes both in time and in frequency domain \cite{SalSchYur,Kri06,Ack07}.  Typically,  the nominal duration of such a pulse is larger than the short time-scale of the 
field fluctuations (i.e., the coherence time).  Such a type of radiation cannot be considered either ergodic or stationary \cite{Good}, and thus simplifications and 
analytic expressions typically used for thermal sources do not apply in the present scenario. For instance, ensemble averages of time-dependent quantities cannot 
be substituted by integrations over time.  In the following we adopt numerical techniques that have 
been developed in the context of quantum optics \cite{FoxPRA88,VanTeiAO80,BilShiPRA90}, in order to produce fluctuating pulses, which exhibit many of  the properties of  SASE-FEL pulses 
in the linear regime. Our algorithm bears analogies to the algorithm used by other authors \cite{RohPRA08,PfeiferOL10}, and the details of our algorithm can be found in \cite{NikLamPRA12}. For the sake of completeness, however, we briefly summarize here its main aspects.

\subsection{Algorithm}
\label{sec2a}
Our algorithm is implemented  on a grid of $N_g$ points in frequency domain, around the central frequency $\omega_s$ of the chosen power spectral density (PSD)  ${\mathscr P}_\zeta(\omega)$.
An independent complex Gaussian random variable $\xi_k$, is assigned to the $k$th point of the grid (of frequency $\omega_k$), with $\aver{\xi_k}=\aver{\xi_k^2}=0$ and $\aver{\xi_k\xi_l^\star}=\delta\omega {\mathscr P}_\zeta(\omega_k)\delta_{k,l}$, where $\delta_{k,l}$ is the Kronecker's delta and $\delta\omega$ is the frequency step on the grid. 
By means of the discrete Fourier transform we generate the colored noise in the time domain  i.e., we  obtain a complex Gaussian random variable $\zeta(t)$ with 
$\aver{\zeta(t)} = 0$ and 
\begin{eqnarray}
\aver{|\zeta(t)|^2}=\int_{-\infty}^\infty d\omega {\mathscr P}_\zeta(\omega).
\label{zeta_var}
\end{eqnarray}
The corresponding autocorrelation function $G_\zeta^{(1)}(t,t^\prime)\equiv \aver{\zeta(t)\zeta^{\star}(t^\prime)}$ is given by
\begin{eqnarray}
G_\zeta^{(1)}(t,t^\prime) = \sum_{k=-N_g/2}^{N_g/2}\delta\omega e^{i\omega_k(t-t^\prime)} {\mathscr P}_\zeta(\omega_k).
\label{cr_zeta}
\end{eqnarray}
This algorithm generates a stationary random noise, and to mimic the fluctuating SASE-FEL pulses one has to be superimpose the noise to a particular Fourier-limited envelope (profile). 

\begin{figure}
\includegraphics[scale=1.0]{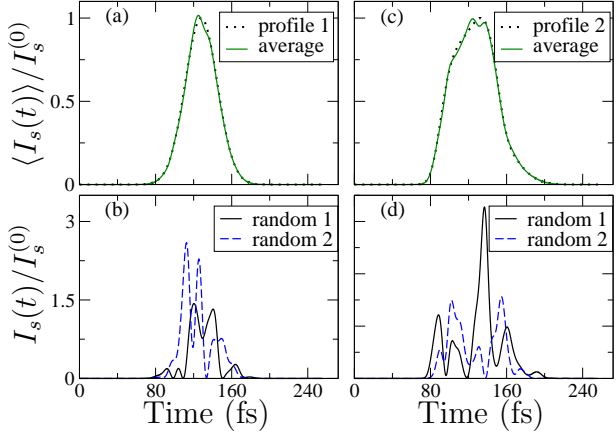}
\caption{(a,c) The dotted curves show two of the profiles $f_s(t)$ used in our simulations.  (b,d) A sample of two random spiky pulses, typically produced in a single realization of the algorithm discussed in Sec. \ref{sec2}, by superimposing Gaussian correlated noise ($\sigma_\omega=0.14$~rad/fs) with the deterministic profiles of (a,c). The solid curves in (a,c) show the average intensity $\aver{I_s(t)}/I_s^{(0)}$ on a sample of 1000 random spiky pulses.}
\label{fig1}
\end{figure}

\subsection{Application}
\label{sec2b}
The above algorithm has been tested for various types of colored noises and various envelopes. In the following we focus on the generation of fluctuating pulses with Gaussian correlated noise, which is the type of correlations typically observed in different SASE FEL facilities \cite{SalSchYur,Kri06,Ack07,Pie96Rei99,VartPRL11,MitzOE08}.  
To this end, the algorithm of Sec. \ref{sec2a}, is seeded with a Gaussian PSD i.e.,  
\begin{eqnarray}
{\mathscr P}_\zeta(\omega)= \frac{1}{\sigma_\omega\sqrt{2\pi}}\exp \left [-\frac{\omega^2}{2\sigma_\omega^2} \right ],
\label{psd}
\end{eqnarray} 
where $\sigma_\omega$ is the standard deviation of the distribution. 

The carrier frequency $\omega_s$ will be included separately later on, while the amplitude of the electric field at $\omega_s$ in a single run of the algorithm 
(within some non-essential multiplicative constants) is defined as  
\begin{eqnarray}
{\cal E}_s(t) = \zeta(t) \sqrt{I_s^{(0)} f_s(t)}, 
\label{Es_t}
\end{eqnarray}
where  $I_s^{(0)} f_s(t)$ is a Fourier-limited (deterministic) pulse profile of finite duration and peak value $I_s^{(0)}$. The intensity of the stochastic pulse in the time domain is simply given by 
\begin{eqnarray}
I_s(t)=|{\cal E}_s(t)|^2=I_s^{(0)}f_s(t)|\zeta(t)|^2.
\label{It}
\end{eqnarray}
The deterministic envelope $f_s(t)$ ensures the smooth rise and drop of the intensity, and can be chosen at will. In Fig. \ref{fig1} we show two of the profiles used for 
$f_s(t)$ throughout our simulations, together with a small sample of spiky pulses. The following discussion will focus on the profile of  Fig. \ref{fig1}(a), which  is a Gaussian  given by 
\begin{eqnarray}
f_s(t) = \exp\left[-\frac{(t-t_0)^2}{\tau^2}\right ],
\label{fst}
\end{eqnarray}
where $\tau$ is the pulse duration and $t_0$ its center. In this case, one can derive analytic expressions for various quantities such as the energy spectral density of the pulses.  
In view of Eqs. (\ref{zeta_var}) and (\ref{It}), averaging over a large number of random pulses one recovers the deterministic pulse $f_s(t)$ i.e., 
\begin{eqnarray}
\aver{I_s(t)}=I_s^{(0)}f_s(t).
\label{It_ft}
\end{eqnarray}
 
 \begin{figure}
	\begin{center}
    \subfigure[]{\label{fig3a}\includegraphics[scale=0.55]{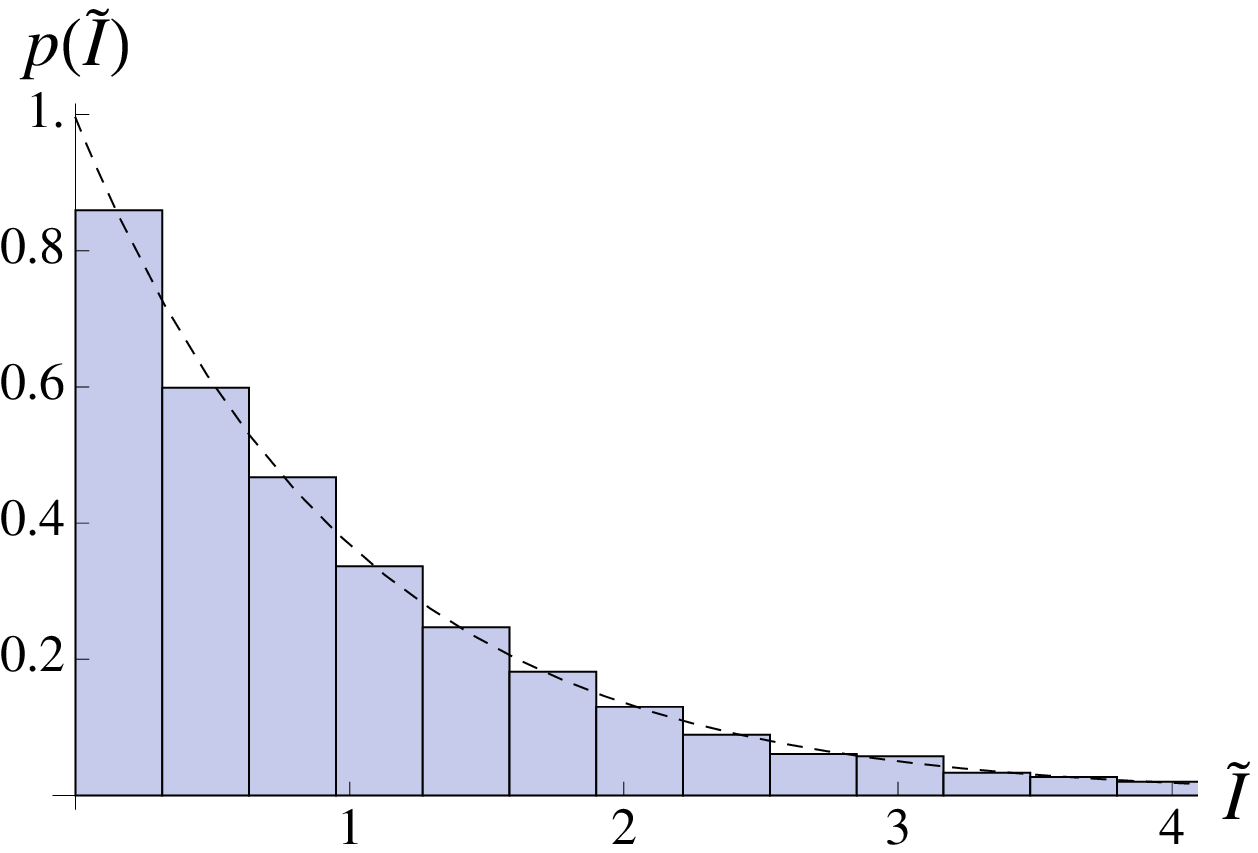}}
    \subfigure[]{\label{fig3b}\includegraphics[scale=0.55]{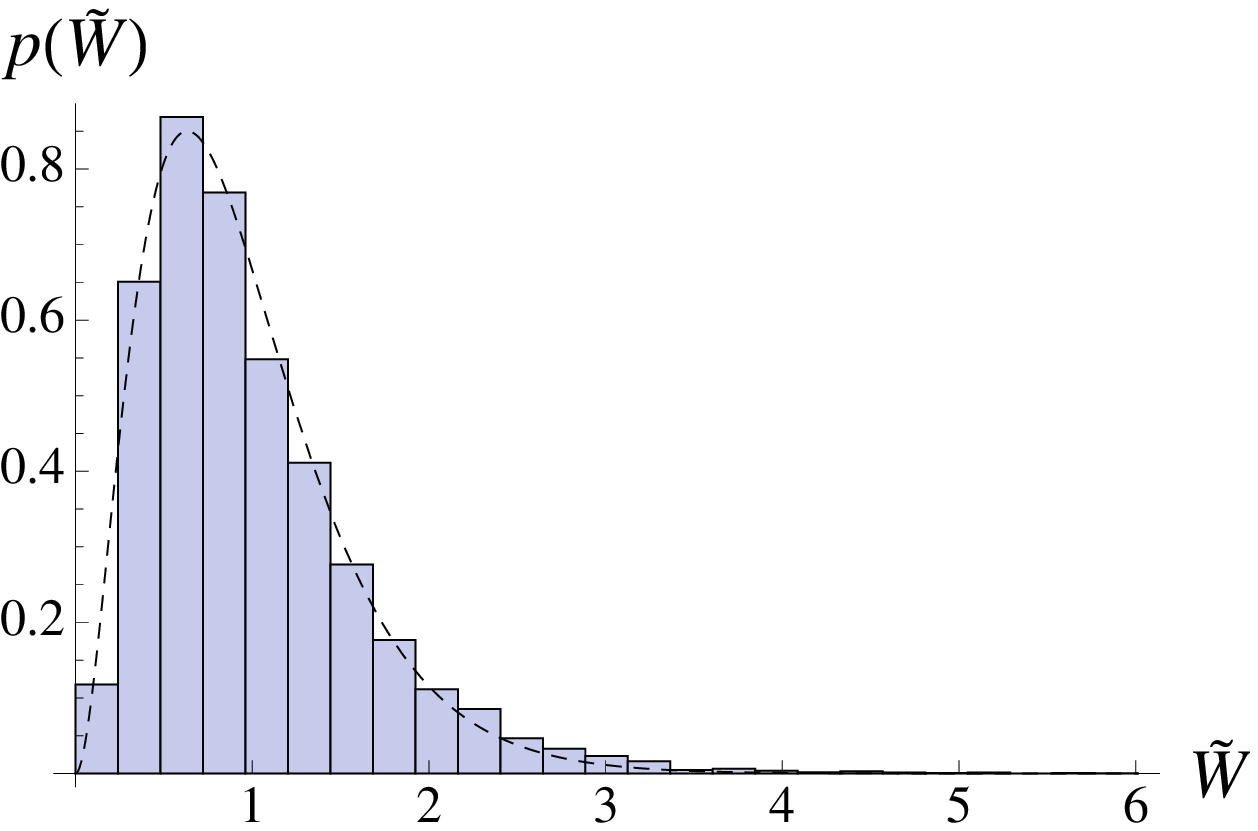}}
  \end{center}
  \caption{Statistics of the generated pulses. Typical  distributions of  (a) the instantaneous intensity $\tilde{I}(t)=I_s(t)/\aver{I_s(t)}$ at $t-t_0=\tau/2$, and (b) 
the energy per pulse $\tilde{W}=W_s/\aver{W_s}$. The dashed lines are fits according to the expected PDFs (see the text). 
Other parameters: $\tau=10$ fs, $\sigma_{\omega}=0.25$ rad/fs, $10^3$ trajectories. }
\label{fig3}
\end{figure}

The fluctuations of the instantaneous electric filed (not shown here) obey a Gaussian distribution. 
In Fig. \ref{fig3a}, we present a sample of the probability distribution of the instantaneous intensity, which is well approximated by the negative exponential probability density function (PDF)
\begin{equation}
p[I_s(t)] = \frac{1}{\aver{I_s(t)}}\exp\left (-\frac{I_s(t)}{\aver{I_s(t)}}\right ).
\label{ned}
\end{equation}
The energy in a random pulse at a space point in the interaction volume is given by  
\begin{equation}
W_s \propto \int_0^\infty I_s(t) d t, 
\end{equation}
and it fluctuates from pulse to pulse. As shown in Fig. \ref{fig3b},  the corresponding probability distribution is well approximated by the Gamma PDF  
\[ 
p(W_s) = \frac{M^MW_s^{M-1}}{\Gamma(M)\aver{W_s}^{M}} \exp\left (-M\frac{W_s}{\aver{W_s}} \right ),  
\label{gd}
\]
where $\Gamma$ here is the Gamma function and $M=\aver{W_s}^2/\aver{(W_s-\aver{W_s})^2}$. The asymptotic forms of $p(W)$ for $M\to 1$ and $M\gg 1$, are the negative exponential and the normal PDFs, respectively. All of these properties are in agreement with experimental observations and theoretical results pertaining to various SASE-FEL facilities \cite{SalSchYur,Kri06,Ack07,Pie96Rei99}.  
 
\begin{figure}
\includegraphics[scale=1.0]{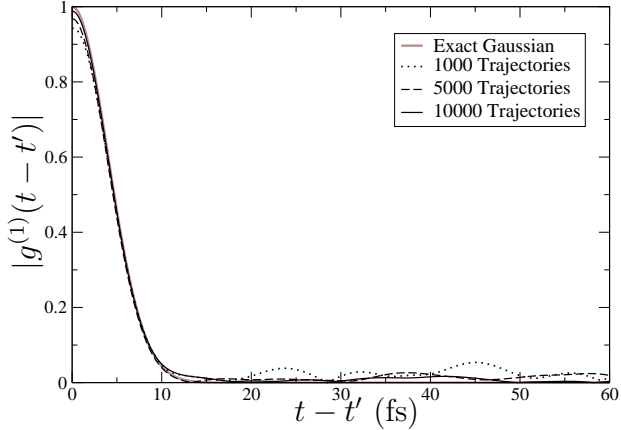}
\caption{Convergence of the algorithm discussed in Sec. \ref{sec2}. The modulus of $g^{(1)}(t,t^\prime)$ is plotted as a function of $t-t^\prime$. Averaging over a large number of pulses (trajectories),  $|g^{(1)}(t,t^\prime)|$ converges to  Eq. (\ref{Gcr}) (grey thick line). Other parameters: $\tau=10$ fs, $\sigma_{\omega}=0.25$ rad/fs.
}
\label{fig4}
\end{figure}

By definition the first-order autocorrelation function is defined as $G^{(1)}(t,t^\prime)=\aver{{\cal E}_s(t){\cal E}_s^\star(t^\prime)}$, and 
using Eqs. (\ref{Es_t}) and (\ref{It_ft}) we obtain
\begin{eqnarray}
|G^{(1)}(t,t^\prime)|&=&\sqrt{\aver{I_s(t)}\aver{I_s(t^\prime)}} |G_\zeta^{(1)}(t,t^\prime)|,
\label{G1}
\end{eqnarray}
where $G_\zeta^{(1)}(t,t^\prime)=\aver{\zeta(t)\zeta^\star(t^\prime)}$. We see therefore that the first-order autocorrelation function for the field not only depends on  
the statistical properties of the noise $\zeta(t)$, but also on the ensemble average of SASE-FEL pulses $\aver{I_s(t)}$. Only in the case of stationary fields one has constant $\aver{I_s(t)}$, and thus $G^{(1)}(t,t^\prime)$ depends solely on the noise correlations and not on the average profiles. In Fig. \ref{fig4} we plot the modulus of the degree of first-order temporal coherence, which is defined as \cite{Loud} 
\begin{eqnarray}
|g^{(1)}(t,t^\prime)|=\frac{|G^{(1)}(t,t^\prime)|}{\sqrt{\aver{I_s(t)}\aver{I_s(t^\prime)}}}. 
\end{eqnarray}
and is equal to $|G_\zeta^{(1)}(t,t^\prime)|$. Clearly, the noise we have generated is Gaussian correlated and the variance of the chosen PSD (\ref{psd}), determines how fast 
the correlations drop with the delay i.e., 
\begin{eqnarray}
|g^{(1)}(t,t^\prime)| = \exp\left [ -\frac{\sigma_{\omega}^2(t-t^\prime)^2}{2}\right ] = |G_\zeta^{(1)}(t,t^\prime)|.
\label{Gcr}
\end{eqnarray}
Some deviations observed for large $t-t^\prime$, can be eliminated by averaging over a  larger number of pulses. 
For a field that obeys Gaussian statistics, Wick's theorem implies that higher-order degrees of coherence can be expressed in terms of $g^{(1)}(t,t^\prime)$, and hence they are determined by the spectral properties of the field \cite{Loud}. 

 Throughout our simulations we have used various types of ${\mathscr P}_\zeta(\omega)$ and thus of correlations, some of which are summarized in table \ref{tab1}, together with the corresponding bandwidths $\gamma$ [i.e., the FWHM of ${\mathscr P}_\zeta(\omega)$], which are multiples of $\sigma_\omega$.
The coherence time is typically defined as 
\begin{eqnarray}
T_c \equiv \int_{-\infty}^{\infty} |g^{(1)}(v)|^2 dv,
\end{eqnarray}
and is also shown in table \ref{tab1}. 

\begin{table}
\caption{\label{tab1} Power spectral densities, bandwidths, and coherence times, for fields with exponential and Gaussian correlations.}
\footnotesize\rm
\begin{tabular*}{\textwidth}{@{}l*{15} {@{\extracolsep{0pt plus12pt} } c} }
\br
Correlations & Power spectral density (PSD) & Bandwidth ($\gamma$) & Coherence time ($T_c$)\\
\mr
Exponential &  $ [\sigma_\omega\pi(\tilde{\omega}^2+1)]^{-1}$  &    $2\sigma_\omega$ & $\sigma_\omega^{-1}$\\ 
Gaussian &   $\exp[-{\tilde{\omega}^2 }/2]/(\sigma_\omega\sqrt{2\pi})$ &   $2\sigma_\omega\sqrt{2\ln(2)}$ & $\sqrt{\pi}\sigma_\omega^{-1}$\\ 
\br
\end{tabular*}
$^{\rm a}$ $\tilde{\omega} = \omega/\sigma_\omega$.
\end{table}

In view of the smooth rise and drop of the intensity for $t\in[0,\infty)$, one can safely assume that $I_s(t)$ is a square integrable function.Thus the energy spectral density of the random pulses is given by ${\mathscr E}_s(\omega)=\aver{|{\cal E}_s(\omega)|^2}$, where ${\cal E}_s(\omega)$ is the Fourier transform of ${\cal E}_s(t)$. Using  Eqs. (\ref{Es_t}),  (\ref{fst}) and (\ref{It_ft}) we find that for
the Gaussian correlated noise of  Eq. (\ref{Gcr}), ${\mathscr E}_s(\omega)$ is also Gaussian. The normalized spectrum $\tilde{\mathscr E}_s(\omega)$, which is obtained by dividing ${\mathscr E}_s(\omega) $ by $\int_{-\infty}^{\infty} {\mathscr E}_s(\omega)d\omega$, is given by 
\bea
\tilde{\mathscr E}_s(\omega) = \frac{2\sqrt{\ln(2)} }{\sqrt{\pi}\Delta\omega_s}\exp\left [-\frac{4\ln(2)\omega^2}{\Delta\omega_s^2}\right ].
\eea
The bandwidth (FWHM) of $\tilde{\mathscr E}_s(\omega)$ is 
\bea
\Delta\omega_s = \Delta\omega_s^{\min}\sqrt{1+\left (\frac{\gamma}{\Delta\omega_s^{\min}}\right )^2}, 
\label{bw2}
\eea
where $\Delta\omega_s^{\min}$ is the bandwidth of the Fourier-limited Gaussian pulse $\aver{I_s(t)}$ of duration $\tau$ and is given by 
\bea
\Delta\omega_s^{\min}  = \frac{2\sqrt{\ln(2)}}{\tau} =  \frac{4\ln(2)}{\Delta t_s}.
\label{fl_bw}
\eea
This is the well-known time-bandwidth relation for Gaussian Fourier-limited pulses \cite{Yariv}, 
with the FWHM of $\aver{I_s(t)}$ denoted by $\Delta t_s = 2\sqrt{\ln(2)}\tau$.  

Equation (\ref{bw2}) shows that for pulses with Gaussian average profile that exhibit Gaussian-correlated fluctuations, the combined 
bandwidth $\Delta\omega_s$ is the geometric mean of the bandwidths corresponding to the Fourier-limited average profile and the fluctuations. For $\gamma\gg \Delta\omega_s^{\min}$, the bandwidth of the pulse is fully determined by the fluctuations i.e., $\Delta\omega_s\simeq\gamma$, 
whereas  for $\gamma\ll \Delta\omega_s^{\min}$, we have the case of a Fourier-limited pulse with $\Delta\omega_s\simeq\Delta\omega_s^{\min}$. The derivation of simple analytic expressions for ${\mathscr E}_s(\omega)$  is a rather difficult task, for non-Gaussian correlated noise and/or for arbitrary average profiles. 

\begin{figure}
\begin{center}
\includegraphics*[scale=1.5]{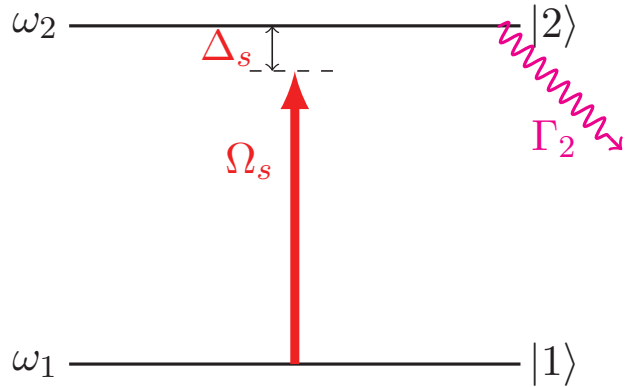}% Here is how to import EPS art
\caption{\label{tls_fig} An atomic transition pertaining to a sub-valence electron, driven by chaotic SASE-FEL pulses. The resulting inner-shell vacancy in the excited atom 
decays with rate $\Gamma_2$.}
\end{center}
\end{figure}

\section{Single Auger resonance}
\label{sec3}
The aforementioned algorithm can be used in studies pertaining to interactions between matter and SASE-FEL radiation with the  statistical properties described above (e.g., see \cite{NikLamPRA12,LamPRA11}), and various types of autocorrelation functions. The present work focuses on the influence of fluctuations on the frequency response of a single Auger resonance. 
Previous studies on the same problem pertained to stationary fields i.e., for time-independent $\aver{I_s(t)}$ \cite{GeoPRA79,ZolPRA79,CamPRA93}, while most of them considered exponentially-correlated fluctuations. Such conditions, however, are not satisfied in the present SASE-FEL facilities, and the ensemble average intensity $\aver{I_s(t)}$ has a finite duration $\tau$. It may be possible to assume some sort of stationarity only if the coherence time $T_c$ is much smaller than the pulse duration $\tau$, and the field is observed for a time window much smaller than $\tau$ (yet much larger than $T_c$). Such a condition may be fulfilled in practise when one monitors directly the light, but it is hard to be fulfilled when matter interacts with SASE-FEL pulses. Typically, in such cases the target (atoms, molecules, etc) experience the rise and fall of each  pulse, and one simply monitors the products of the interaction (i.e., electrons, ions, etc). In the end, the experimental data to be 
interpreted consist of averages over many pulses. Under such circumstances, the SASE-FEL radiation may or may not be considered stationary and any theoretical description has to take into account the finite temporal width of $\aver{I_s(t)}$.  As long as $\aver{I_s(t)}$ is a smooth function, its details besides the duration $\tau$, are not expected to play a significant role in most cases. For the problem under consideration, we have confirmed this fact by considering different profiles for $\aver{I_s(t)}$, two of which are shown in Figs. \ref{fig1}(a) and (c). 

In the following we consider the case of a sub-valence atomic transition depicted in Fig. \ref{tls_fig}. It is driven by a SASE-FEL beam that is focused on a target of neutral atoms, inducing an electric-dipole transition of an inner-shell electron from state $\ket{1}$ to a highly excited state $\ket{2}$. The relaxation of the vacancy in the excited atom via autoionization, gives rise to resonant-Auger (RA) electrons which are observed in the experiment.  The difference between the resonant Auger considered in this paper from what is referred to as normal Auger should perhaps be noted here, before embarking on the formal treatment. Resonant Auger involves the  excitation of a core electron to a discrete state of the neutral, which means that the exciting radiation can be tuned around the corresponding resonant frequency. In the normal Auger the core electron is photoionized in a bound-continuum transition, in which case there is no resonant frequency to tune around. In both cases it is the electrons ejected through autoionization in the filling of the hole. The width of the Auger electrons' energy spectrum observed in both cases, reflects the lifetime of the core-hole state. Our concern in this paper is the dependence of the observed line-shape for the RA electrons on the field fluctuations, as the radiation is tuned around the resonant transition. Clearly, this does not have a counterpart in normal Auger. Of course another difference between the two types of processes is that normal Auger leads to a doubly charged ion, which has no bearing on our calculations.

Let $\hbar\omega_j$ denote the energy of the state $\ket{j}$, and let $\Gamma_2$ be the rate associated with the  decay. 
The electric field of the radiation is
\[{\bf E}_s(t)=\left [{\cal E}_s(t)e^{i\omega_st}+{\cal E}_s^\star (t)e^{-i\omega_st}\right ] {\bf e}_s,\]
where $\omega_{s}$ denotes the central frequency of the spectrum, ${\bf e}_s$ is the polarization vector, and ${\cal E}_s(t)$ is the fluctuating complex amplitude. 
Throughout this work ${\cal E}_s(t)$ is treated as a stochastic complex Gaussian random function and is generated along the lines of the previous section. 
The instantaneous Rabi frequency $\Omega_{s}(t)$, given by 
\begin{eqnarray}
\Omega_{s}(t)=\frac{\vec{\mu}_{12}\cdot{\bf e}_s {\cal E}_s(t)}{\hbar},
\label{stocOmega0}
\end{eqnarray} 
is also a stochastic complex Gaussian random variable with zero mean and variance determined by the 
variance of the field. In this definition, $\hat{\vec{\mu}}$ is the electric dipole operator, and $\vec{\mu}_{12}$ is the transition dipole moment for $\ket{1}\leftrightarrow\ket{2}$. In the following for the sake of brevity we also write $\mu_{12}=\vec{\mu}_{12}\cdot{\bf e}_{s}$. 
In view of Eqs. (\ref{Es_t}) and (\ref{It_ft}), we have
\begin{eqnarray}
\Omega_s(t) = \frac{\mu_{12}\sqrt{\aver{I_s(t)}}}{\hbar}\zeta(t)= \Omega_s^{(0)}  \sqrt{f_s(t)} \zeta(t).
\label{stocOmega1}
\end{eqnarray}
where 
\begin{eqnarray}
\Omega_s^{(0)} =\frac{\mu_{12}\sqrt{I_s^{(0)}}}{\hbar}
\label{stocOmega2}
\end{eqnarray}
is the peak value of the Rabi frequency. 

The problem can be formulated in the framework of the reduced atomic density matrix with elements $\rho_{ij}(t)$. 
In the rotating-wave approximation, the equations of motion for $\rho_{ij}$ read  
\begin{eqnarray}
&&\frac{\partial\sigma_{11}}{\partial t}=2{\rm Im}[\Omega_s^\star\sigma_{12}]
\label{tla_ode1}\\
&&\frac{\partial\sigma_{22}}{\partial t}=-\Gamma_2\sigma_{22}-2{\rm Im}\left [\Omega_s^\star\sigma_{12} \right ]\label{tla_ode2}\\
&&\frac{\partial\sigma_{12}}{\partial t}=\left (i\Delta_s-\frac{\Gamma_{21}}{2}\right )\sigma_{12}+i\Omega_s(\sigma_{22}-\sigma_{11}),\label{tla_ode3}
\end{eqnarray}
where $\rho_{ii}=\sigma_{ii}$, $\rho_{12}=\sigma_{12}e^{i\omega_s t}$, $\Delta_s=\omega_{21}-\omega_{s}$ is the detuning of the field from resonance, while Stark shifts have been neglected. In the absence of other types of (in)homogeneous broadening mechanisms, we have $\Gamma_{21}=\Gamma_2$. 

The frequency response of the atoms to the SASE-FEL radiation, as we vary $\omega_s$ around resonance, is obtained by monitoring the RA electrons. The total probability for decay or else the yield, in terms of the fraction of the initial neutral population, is given by 
\begin{eqnarray}
Q_2 = \Gamma_2\int_0^{\infty} dt \sigma_{22}(t).
\label{q2_eqa}
\end{eqnarray}
Alternatively, we can add to the equations of motion for $\sigma_{ij}$, the following differential equation 
\begin{eqnarray}
\frac{\partial Q_2}{\partial t}=\Gamma_2\sigma_{22},
\label{q2_eqb}
\end{eqnarray}
where $\Gamma_2$ is the probability per unit time for  decay. 

In the presence of fluctuations in the electric field,  equations (\ref{tla_ode1})-(\ref{tla_ode3}) and (\ref{q2_eqb}) constitute a set of coupled stochastic differential equations. 
Our simulations involve many trajectories, and in each one of them a SASE-FEL pulse ${\cal E}_s(t)$ is generated randomly according to the algorithm described in Sec. \ref{sec2}. The 
stochastic differential equations are propagated from $t=0$ to $t=T_f$, where $T_f\gg\tau$. This is essentially equivalent to taking the upper limit of Eq. (\ref{q2_eqa}) to infinity. 
The average stochastic signal $\aver{Q_2}$ is obtained by averaging over a large number of random pulses (trajectories). 

The scheme of Fig. \ref{tls_fig}, has been at the core of many experiments at various FEL facilities. Our results will be discussed in connection with the recent experiment by Mazza {\em et al.} \cite{MazJPB12}, pertaining to the the spectral response of the  Auger resonance 3d$\to$5p in Kr.  To keep our formalism as general as possible, in the following discussion the various quantities are measured in units of the natural linewidth $\Gamma_2$,  which for the Auger resonance 3d$\to$5p in Kr is $83$ meV, and corresponds to a lifetime of about $8$ fs \cite{MazJPB12}. 

\subsection{Decorrelation of atom-field dynamics: A route to analytical solutions}
\label{sec3a}
Integrating formally the equation for $\sigma_{12}$, substituting the result into the equation for $\sigma_{22}$, and taking the stochastic average we obtain
\begin{eqnarray}
\frac{\partial\aver{\sigma_{22}}}{\partial t}&=& -\Gamma_2\aver{\sigma_{22}}-2{\rm Re}\left [\lambda(t) \right ]
\label{s22}
\end{eqnarray}
where  
\begin{eqnarray}
\lambda(t)\equiv  \int_0^{t} dt^{\prime} e^{\alpha (t-t^{\prime})} \aver{\Omega_s^\star(t)\Omega_s(t^{\prime})n(t^{\prime})},
\label{lambda1}
\end{eqnarray}
with $\alpha=i\Delta_s-\Gamma_{21}/2$, and the population difference given by $n(t) \equiv \sigma_{22}(t)-\sigma_{11}(t)$. 
We see that the dynamics of the system are determined by the atom-field correlation function $\aver{\Omega_s^\star(t)\Omega_s(t^{\prime})n(t^{\prime})}$. In order to proceed 
further analytically, one has to decorrelate this function according to 
\begin{eqnarray}
\label{decor}
\aver{\Omega_s^\star(t)\Omega_s(t^{\prime})n(t^{\prime})}\approx \aver{\Omega_s^\star(t)\Omega_s(t^{\prime})}\aver{n(t^{\prime})}.
\end{eqnarray}
In view of Eq. (\ref{stocOmega0}) one then obtains 
\begin{eqnarray}
\lambda(t) &=& \frac{|\mu_{12}|^2}{\hbar^2}  \int_0^{t} dt^{\prime} e^{\alpha (t-t^{\prime})} G^{(1)}(t,t^\prime)\aver{n(t^{\prime})}.
%&=&|\Omega_s^{(0)}|^2\int_0^{t} dt^{\prime} e^{\alpha (t-t^{\prime})} \sqrt{f_s(t)f_s(t^{\prime})} G_\zeta^{(1)}(t, t^{\prime}) \aver{n(t^{\prime})}.\nonumber
\label{lambda2}
\end{eqnarray}
The decorrelation therefore enables one to express $\lambda(t)$ in terms of the autocorrelation function of the field $G^{(1)}(t,t^\prime)$. For exponentially correlated fields i.e., for
\begin{eqnarray}
G^{(1)}(t,t^\prime) = \sqrt{\aver{I_s(t)}\aver{I_s(t^\prime)}}e^{-\gamma|t-t^\prime|/2},
\label{exp_cor}
\end{eqnarray}
one has 
\begin{eqnarray}
\lambda(t) = \frac{|\mu_{12}|^2}{\hbar^2}  \int_0^{t} dt^{\prime} e^{\tilde{\alpha} (t-t^{\prime})} \sqrt{\aver{I_s(t)}\aver{I_s(t^\prime)}} \aver{n(t^{\prime})}
\label{lamb}
\end{eqnarray}
where the bandwidth $\gamma$ of the Lorentzian PSD of the noise, has been absorbed in $\tilde{\alpha}$ i.e., $\tilde{\alpha} = \alpha-\gamma/2 = i\Delta_s-(\Gamma_{21}+\gamma)/2$.  
This shows that when the decorrelation is valid, the average dynamics of the two-level system (TLS) driven by a stochastic field with exponential correlations 
can be basically obtained from the dynamics of the TLS driven by a Fourier-limited  pulse of the same average intensity $\aver{I_s(t)}$, by setting $\Gamma_{21}\to \Gamma_{21}+\gamma$ [hereafter, for the reasons explained below, this model is referred to as the phase-diffusion (PDM) model]. This is a well known result in the case of stationary fields \cite{AgaPRA78,GeoPRA78,GeoPRA79}, and here we show that it  holds for non-stationary fields as well. 
However, it has to be emphasized that it is intimately connected to the exponential form of the correlation function (\ref{exp_cor}). For fields that are not exponentially correlated, the analytic treatment of 
the problem even under the decorrelation approximation is a rather cumbersome (if not impossible) task, unless one employees additional approximations. For instance, in the case of  weak fields and for sufficiently short times, we can further assume that the atomic populations do not vary significantly during the pulse, setting $n(t)\approx -1$ in Eq. (\ref{lambda2}). In this case, $\lambda(t)$ is determined solely by the statistical properties of the field, as well as the ensemble average intensity profile $\aver{I_s(t)}$, allowing thus in principle for an analytic treatment for special cases of $\aver{I_s(t)}$  and $G_\zeta^{(1)}(t,t^\prime)$. 

\begin{figure}
\begin{center}
\includegraphics[scale=1.5]{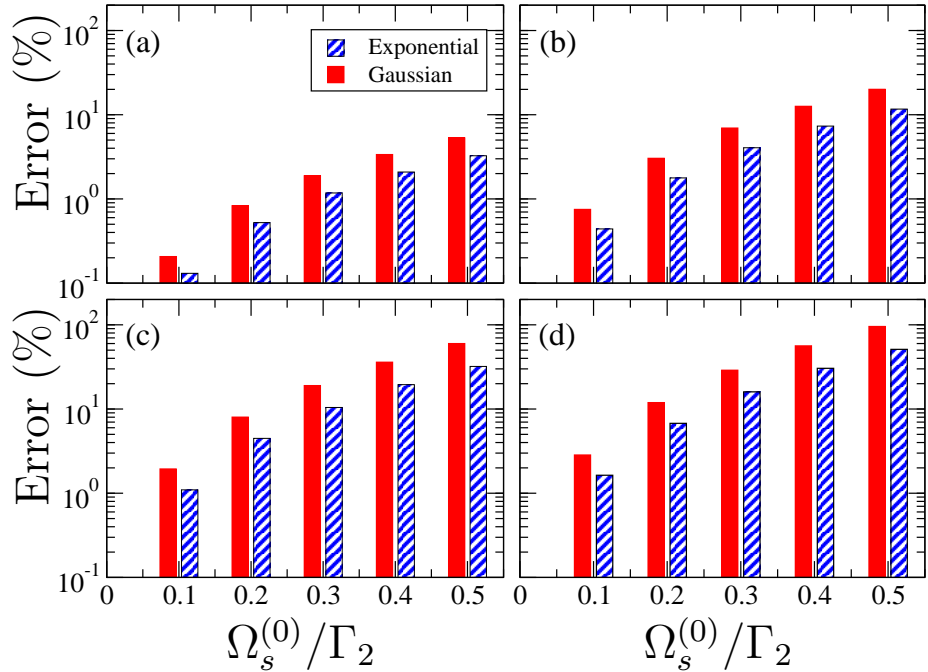}% Here is how to import EPS art
\caption{\label{decor:fig} The relative error (\ref{error}), for Gaussian and exponentially correlated noises, as a function of $\Omega_s^{(0)}/\Gamma_2$, for various bandwidths of the field. (a) $\gamma=6.67\Gamma_2$; (b) $\gamma=3.33\Gamma_2$; (c) $\gamma=1.67\Gamma_2$; (d) $\gamma=1.11\Gamma_2$. 
Other parameters: Gaussian pulse profile, $\Gamma_2\tau=3$, 5000 random pulses, $\Delta_s=0$, $t=t_0$.}
\end{center}
\end{figure}

It has been shown rigorously \cite{AgaPRA78,GeoPRA78,GeoPRA79}, that the decorrelation (\ref{decor}) is valid for fields that satisfy 
\[\aver{{\cal E}_s^\star(t_1){\cal E}_s(t_2)\cdots {\cal E}_s^\star(t_{2m-1}){\cal E}_s(t_{2m})} = \prod_{j}\aver{{\cal E}_s^\star(t_j){\cal E}_s(t_{j+1})},\]
where the product is for odd $j\in\{1,\ldots,2m-1\}$. This is, for instance, the case of a phase-diffusion field i.e., a field with constant amplitude and Wiener-Levy statistics for its phase.  
As discussed in Sec. \ref{sec2}, however, for a SASE FEL operating in the linear regime, the field has the statistical properties of a chaotic field (with amplitude and phase fluctuations), 
and as such cannot satisfy this relation. It has been conjectured that for general stochastic fields, the decorrelation is expected to be a good approximation as long as the  field fluctuations are much faster than any variations in the atomic populations \cite{AgoJPB78}. Formally speaking, for the TLS under consideration the coherence time $T_c$ has to be much smaller than all of the other characteristic 
time scales that are associated with the atomic populations  i.e.,  
\bea
\gamma\sim T_c^{-1} \gg \max\{\Omega_s^{(0)}, \Gamma_2\}. 
\label{cond}
\eea
Our model enabled us to check this conjecture in the context of  Gaussian and  exponentially correlated noises. According to the above equations, the decorralation affects the populations through the real part of $\lambda(t)$, and the associated errors can be quantified by 
\bea
\textrm{Error}(t) = \frac{\left | \lambda(t) - \tilde{\lambda}(t)\right |}{\left | \lambda(t)\right|}\times100\%,
\label{error}
\eea
where $\tilde{\lambda}(t)$ is given by Eq. (\ref{lambda1}) with the decorrelation (\ref{decor}).
%\bea
%\fl \textrm{Error}(t) = \left |\int_0^{t} dt^{\prime} e^{\alpha (t-t^{\prime})} \aver{\Omega_s^\star(t)\Omega_s(t^{\prime})n(t^{\prime})} - 
%\int_0^{t} dt^{\prime} e^{\alpha (t-t^{\prime})} \aver{\Omega_s^\star(t)\Omega_s(t^{\prime})}\aver{n(t^{\prime})} \right |,
%\eea
We calculated this quantity numerically for various values of $\gamma$, $\Omega_s^{(0)}$ and $t$, and in Fig. \ref{decor:fig} we show the behaviour of the errors for $t=t_0$ (the depicted behavior is analogous for $t\neq t_0$). 
Clearly, for a fixed $\gamma$  the decorrelation is well justified  for weak $\Omega_s^{(0)}$, whereas it starts breaking down (the errors increase considerably) as we  increase the ratio $\Omega_s^{(0)}/\Gamma_2$. For larger $\gamma$, the errors remain small even for moderate values of  $\Omega_s^{(0)}/\Gamma_2$ [e.g., see Fig. \ref{decor:fig}(a)] whereas for smaller values of $\gamma$, the decorrelation  errors are rather low only for $\Omega_s^{(0)}\lesssim 0.1\Gamma_2$ [e.g., see Fig. \ref{decor:fig}(c,d)]. In other words, 
for fixed ratio $\Omega_s^{(0)}/\Gamma_2$, the decorrelation errors increase with decreasing $\gamma$. Analogous observations 
are expected for $\Delta_s\neq 0$, albeit the same level of errors are expected to occur at larger peak Rabi frequencies. 

In view of these results, and given that many SASE FEL experiments are currently performed in the regime of weak fields (i.e., $\Omega_s^{(0)}<0.1\Gamma_2$), a question arises here is whether various experimental observations can be described 
in the framework of a PDM with decorrelated atom-field dynamics. This would facilitate considerably the theoretical analysis 
of experimental results, since one does not have to take ensemble averages over many randomly fluctuating pulses. In the following section we address this question for a particular observable, namely the total yield of RA electrons in the process of Fig. \ref{tls_fig}. 

\begin{figure}
\begin{center}
\includegraphics[scale=1.5]{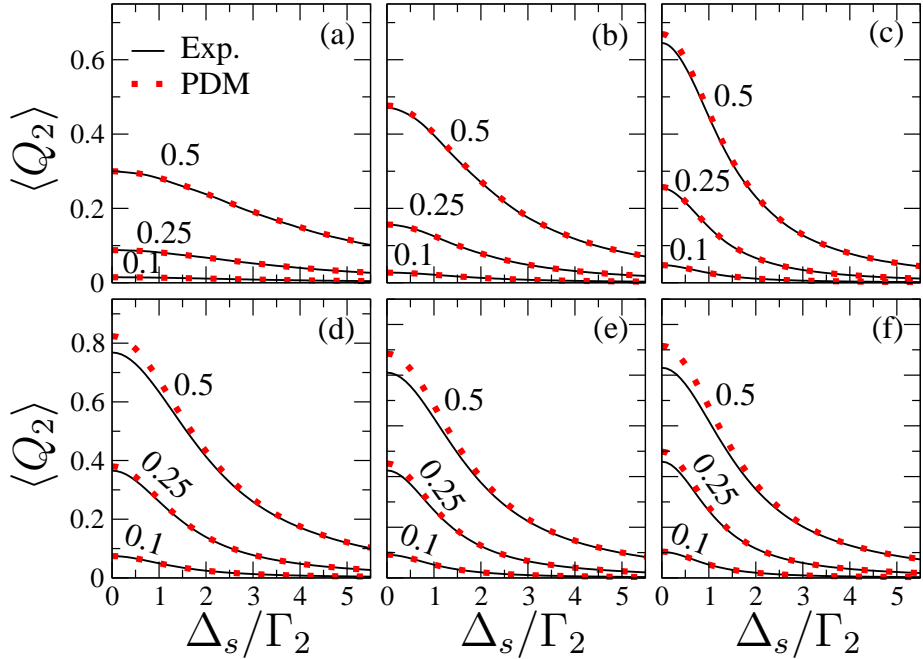}% Here is how to import EPS art
\caption{\label{line1} Single resonance driven by stochastic pulses with phase fluctuations only (PDM), and phase+amplitude exponentially correlated fluctuations. The average total yield of RA electrons $\aver{Q_2}$ is plotted as a function of the detuning of the field from resonance, for three 
different values of the ratio $\Omega_s^{(0)}/\Gamma_2$, and for various bandwidths of the field. (a) $\gamma=13.33\Gamma_2$;  (b) $\gamma=6.67\Gamma_2$; (c) $\gamma=3.33\Gamma_2$; (d) $\gamma=1.67\Gamma_2$; (e) $\gamma=1.11\Gamma_2$; (f) $\gamma=0.83\Gamma_2$. Other parameters: Gaussian pulse profile, $\Gamma_2\tau=3$, 2000 random pulses. The signal is symmetric with respect to $\Delta_s=0$, and only the part for positive $\Delta_s$ is shown.}
\end{center}
\end{figure}

%%%%%%%
% TLA - Numerical Results
%%%%%%%

\subsection{Effects of SASE-FEL field fluctuations on the total yield of resonant-Auger electrons}
\label{sec3b}

In general, the dependence of the total yield  $Q_2$ on the detuning $\Delta_s$  conveys information about the natural linewidth of the observed resonance, as well as the strength and the bandwidth of the driving field. Recent experiments in various FEL facilities pertain to weak (e.g., see \cite{MazJPB12}) as well as strong driving of resonances (e.g., see \cite{KanterPRL11}).
As discussed in \cite{NikLamPRA12}, in the case of strong or even moderate fields (i.e., for $\Omega_s^{(0)}\gtrsim \Gamma_2$), 
$Q_2(\Delta_s)$ may exhibit unconventional lineshapes, as a result of power broadening. In contrast to what is typically found in standard textbooks, in the framework of SASE FEL pulses the driving of an atomic transition is pulsed and the broadening thus depends on both the peak intensity and the duration of the pulse. Throughout this section we will focus on the case of weak fields (i.e., for $\Omega_s^{(0)}\ll \Gamma_2$), which are currently accessible to various SASE FEL facilities. For instance, the intensities reported in the recent experiment by Mazza {\em et al.} \cite{MazJPB12} are in the range of about $10^{11}-10^{12} \rm{W\,cm}^{-2}$, which means that the peak Rabi frequencies experienced by the atoms were at least three orders of magnitude smaller than the natural linewidth of the Auger resonance 3d$\to$5p in Kr. Moreover, the reported pulse durations (FWHM) were a few tens of $\Gamma_2$.  

\begin{figure}
\begin{center}
\includegraphics[scale=1.5]{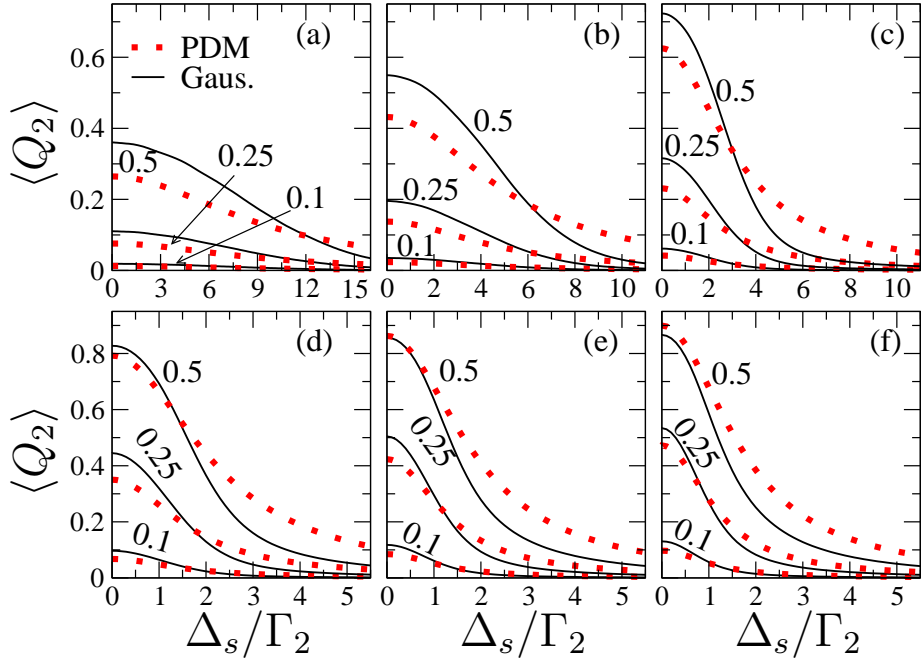}% Here is how to import EPS art
\caption{\label{line2} As in Fig. \ref{line1}, for phase+amplitude Gaussian correlated fluctuations and bandwidths: 
(a) $\gamma=15.70\Gamma_2$;  (b) $\gamma=7.85\Gamma_2$; (c) $\gamma=3.92\Gamma_2$; (d) $\gamma=1.96\Gamma_2$; (e) $\gamma=1.31\Gamma_2$; (f) $\gamma=0.98\Gamma_2$.}
\end{center}
\end{figure}

The theoretical model adopted  in \cite{MazJPB12} for the interpretation of the  experimental observations is basically the PDM. 
As depicted in Fig. \ref{line1}, the line-shape  $Q_2(\Delta_s)$ within the PDM is in a very good agreement with the line-shape obtained for an exponentially correlated field with both amplitude and phase fluctuations, apart perhaps from small deviations around $\Delta_s=0$ for small values of $\gamma$ and moderate values of $\Omega_s^{(0)}$. This is in agreement with the previous discussion on Fig. \ref{decor:fig}. The situation is substantially different when the predictions of the PDM are compared to a Gaussian correlated field with both amplitude and phase fluctuations. As shown in Fig. \ref{line2}, the two lineshapes deviate considerably for $\Omega_s^{(0)}<\Gamma_2$. By contrast to the case of exponentially-correlated fields, 
and despite the weak driving, the lineshapes within the PDM exhibit lower peaks and are broader (they drop much slower) than the ones in our simulations for fields with Gaussian-correlated amplitude and phase fluctuations. As mentioned above, the PDM refers to a field with constant amplitude and exponentially-correlated phase fluctuations. Hence, these discrepancies can be attributed to the differences in the nature of the underlying Lorentzian and Gaussian PSDs.

\begin{figure}
\begin{center}
\includegraphics[scale=1.5]{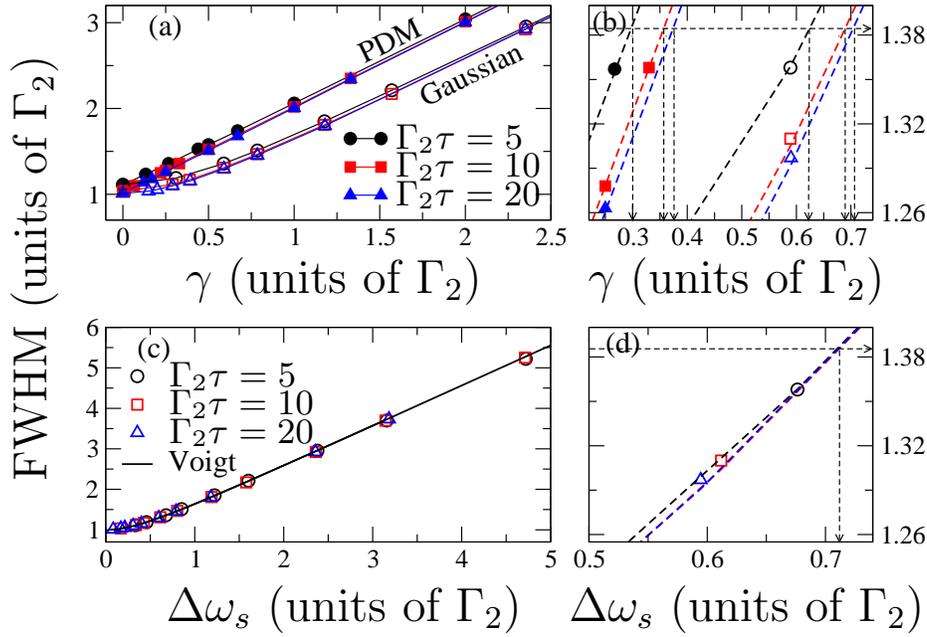}% Here is how to import EPS art
\caption{\label{fw_fig} Single resonance driven by stochastic pulses. The FWHM of the total yield of RA electrons $\aver{Q_2}$ is plotted as a function of the field bandwidth.  
(a) The dependence of the FWHM on the bandwidth $\gamma$ (see table \ref{tab1}), for the PDM (filled symbols) and the Gaussian-correlated noise (open symbols), for various pulse durations.  
(b) A close up of Fig. \ref{fw_fig} (a) for the FWHM reported in \cite{MazJPB12}, i.e., $\approx 1.38\Gamma_2$. The vertical dashed arrows mark the corresponding values of $\gamma$ for the various models and pulse durations. 
The thick dashed curves correspond to the best fits to the numerical data (symbols).
(c) The dependence of the FWHM on the combined bandwidth $\Delta\omega_s$, for Gaussian-correlated noise and various pulse durations.  The solid line is the FWHM corresponding 
to the Voigt profile [see Eq. (\ref{VoigtWidth})]. (d) A close up of Fig. \ref{fw_fig} (c) for the FWHM reported in \cite{MazJPB12}. The vertical dashed arrow marks the corresponding value of $\Delta\omega_s$. The thick  dashed curves correspond to the best fits to the numerical data, and they are very close  to the solid curve of Fig. \ref{fw_fig} (c). Other parameters: Gaussian pulse profile, $\Omega_s^{(0)}=10^{-2}\Gamma_2$, 2000 random pulses.}
\end{center}
\end{figure}

Let us consider for instance the FWHM (linewidth) of the total yields of RA electrons in the two cases.  In the absence of fluctuations, i.e., for Fourier-limited pulses, 
the FWHM is expected to be equal to the natural linewidth $\Gamma_2$ only in the limit of  very weak $\Omega_s^{(0)}\ll\Gamma_2$, and  
very long pulses i.e., for $\Delta\omega_s^{\min}\ll\Gamma_2$. In the experiment \cite{MazJPB12}, both of these conditions were satisfied and thus the excess observed FWHM $(\approx 1.38\Gamma_2)$ can be attributed solely to the presence of fluctuations in the SASE FEL pulses. Given that a PDM was employed for the theoretical interpretation of these observations, we have compared the predictions of this model for the parameters of the experiment, relative to a numerical simulation that takes a statistical average over many spiky pulses with Gaussian-correlated phase and amplitude fluctuations. In Fig. \ref{fw_fig}(a) we present the predictions of the two models for the dependence of the FWHM of the total yield of RA electrons on the bandwidth of the field.  As was expected, for both models the FWHM approaches $\Gamma_2$ for very small $\gamma$, and increases with increasing $\gamma$. The FWHM that corresponds to  a given bandwidth  depends only weakly on the nominal pulse duration $\tau$, since $\Delta\omega_s^{\min}\ll\gamma$ [see Eq. (\ref{bw2})]. 
The main observation, however, is that the PDM  predicts a linear increase of the FWHM with $\gamma$, whereas  in the simulations with Gaussian-correlated noise one obtains a nonlinear increase. This is a crucial difference that has to be taken into account when deducing the bandwidth of the field, by observing the deviations of the average  linewidth of $\aver{Q_2(\Delta_s)}$ from $\Gamma_2$. Typically, the bandwidth  that corresponds to a particular FWHM within the  PDM  is always smaller than the one for Gaussian-correlated fields.  For instance, the bandwidths that correspond to the FWHM reported in \cite{MazJPB12},  typically differ by almost a factor of 2  [see Fig. \ref{fw_fig}(b)], for all the depicted pulse durations.

\begin{figure}
\begin{center}
\includegraphics[scale=1.5]{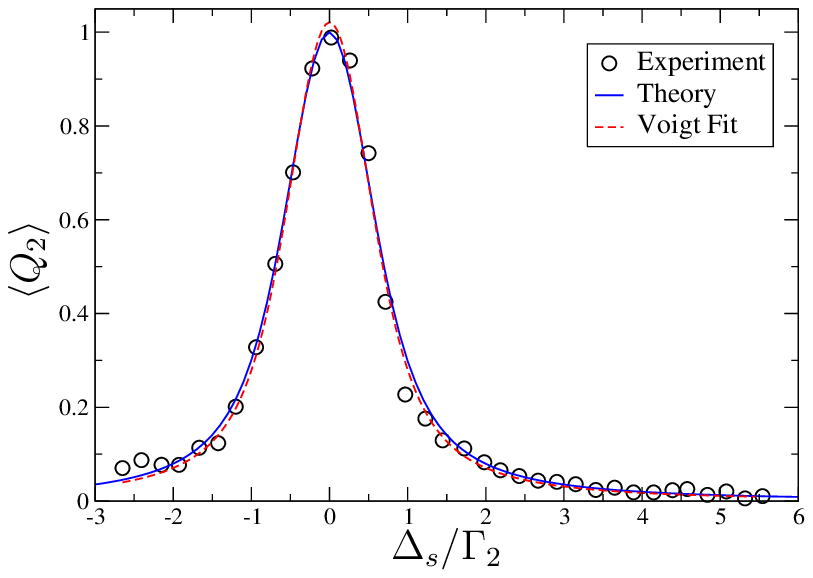}% Here is how to import EPS art
\caption{\label{exp_fig} Single resonance driven by stochastic pulses. The average total yield of RA electrons $\aver{Q_2}$ is plotted as a function of the detuning of the driving field from resonance.  
The symbols are experimental data that have been obtained from the work of  \cite{MazJPB12}, and the solid line shows the average signal obtained from our simulations with Gaussian correlated field. The dashed line is a fit to the experimental data based on the Voigt profile.
Other parameters: Gaussian pulse profile, $\gamma=0.72\Gamma_2$, $\Gamma_2\tau=20$, $\Omega_s^{(0)}=10^{-2}\Gamma_2$, 2000 random pulses.}
\end{center}
\end{figure}

Let us focus now on the simulations with Gaussian-correlated noise. When plotting the FWHM of the average total yield of RA electrons as a function of the combined bandwidth $\Delta\omega_s$ [see Fig. \ref{fw_fig}(c)], one finds that the numerical data (irrespective of the nominal pulse duration), are well approximated  by the same expression, namely
\bea
\Gamma_{\rm voigt} \approx \Gamma_2\left [
0.5346 +\sqrt{0.2166+\left (\frac{\Delta\omega_s}{\Gamma_2}\right )^2}\,
\right ].
\label{VoigtWidth}
\eea
This is the linewidth of a Voigt profile that stems from the combination of a Lorentzian (with FWHM $\Gamma_2$) and a Gaussian (with FWHM $\Delta\omega_s$). The bandwidth of the field that corresponds to the FWHM reported in \cite{MazJPB12}, is approximately 
$0.72\Gamma_2$, which is in very good agreement with the experimental estimations [see Fig. \ref{fw_fig}(d)]. Our model, is also able to reproduce not only the FWHM, but rather the entire experimentally  observed lineshape of the average total yield of RA electrons. The data points (circles) in Fig. \ref{exp_fig} have been extracted from Fig. 3 of  \cite{MazJPB12}, whereas the solid curve is the average total yield obtained by propagating numerically the differential Eqs. (\ref{tla_ode1})-(\ref{q2_eqb}), 
and averaging over of many stochastic pulses. The chosen nominal pulse duration and the peak Rabi frequency are within the range  
of values reported in \cite{MazJPB12}. Clearly, there is an excellent agreement between the theoretical curve and the experimental data. 
In the same figure we also show a fit to the experimental data based on the Voigt profile (dashed curve). This fit corresponds to $\Gamma_{\rm voigt}\simeq 1.34\Gamma_2$, and is slightly worse than what is obtained from our simulations.

\section{Conclusions}
We have studied the frequency response of a weakly driven two-level system, in the presence of SASE FEL field with amplitude and phase fluctuations. 
By contrast to \cite{RohPRA08,KanterPRL11,KazanskyPubs}, we have not analysed the actual spectrum of the resonant-Auger electrons, but rather we focused on the dependence of the total  yield of resonant-Auger electrons on the detuning of the driving field from resonance; a less demanding quantity that can be measured in related experiments.
 It has been shown that for exponentially correlated fluctuations, the lineshape of the average total yield of resonant-Auger electrons is well described by a simple phase-diffusion model with phase-fluctuations only, and decorrelated atom-field dynamics. On the contrary, the same model fails to capture all of the effects of fields with Gaussian-correlated fluctuations. In particular, even in the regime where the decorrelation of atom-field dynamics is valid, the PDM predicts a linear increase of the linewidth with the bandwidth of the field, whereas the increase for Gaussian-correlated fluctuations is nonlinear and well-approximated by an expression corresponding to a Voigt profile. Our results have been also discussed in connection with recent experimental data from FLASH, and an excellent agreement has been found.

\end{document}